\documentclass[10pt,letterpaper,osajnl2,preprint,showpacs,superscriptaddress]{revtex4}
\usepackage[draft]{hyperref} 
\usepackage{mathrsfs}
\usepackage{graphicx}
\usepackage{setspace}

\begin{document}

\title{Forming positive-negative images using conditioned partial measurements from reference arm in ghost imaging}

\author{Jianming Wen$^{1,2}$}
\affiliation{$^1$National Laboratory of Solid State Microstructures, School of Physics, School of Engineering and Applied Science, Nanjing University, Nanjing 210093, China\\
$^2$Institute for Quantum Information Science \& Department of Physics and Astronomy, University of Calgary, Calgary, Alberta T2N 1N4, Canada}
\email{jianming.wen@gmail.com}

\date{\today}

\begin{abstract} A recent thermal ghost imaging experiment implemented in the Wu's group [Chin. Phys. Lett. \textbf{279}, 074216 (2012)] showed that both positive and negative images can be constructed by applying a novel algorithm. This algorithm allows to form the images with use of partial measurements from the reference arm, even which never passes through the object, conditioned on the object arm. In this paper, we present a simple theory which explains the experimental observation, and provides an in-depth understanding of conventional ghost imaging. In particular, we theoretically show that the visibility of formed images through such an algorithm is not bounded by the standard value $\frac{1}{3}$. In fact, it can ideally grow up to unity (with reduced imaging quality). Thus, the algorithm described here not only offers an alternative way to decode spatial correlation of thermal light, but also mimics a ``bandpass filter" to remove the constant background such that the visibility or imaging contrast is improved. We further show that conditioned on one still object present in the test arm, it is possible to construct its image by sampling the available reference data.
\end{abstract}

\maketitle

\section{Introduction}
In classical optics, the spatial distribution of a physical object is estimated through the imaging process by measuring the emitted optical radiation, or by making use of an optical wave that interacts with the object, via transmission or reflection. An extended detector such as a CCD camera or an array detector is usually applied to measure the spatial distribution of the optical intensity. In an interferometric system, the spatial distribution of the optical field is inferred from measurements of the light intensity \cite{teich}. The emergence of coherence theory \cite{glauber,mandel} in 1960s spurred the development of new type of imaging systems based on measurements of the second-order correlation function (i.e., measuring intensity correlation or the photon coincidence counts) at pairs of points in the detection plane. A well-known example of imaging an object emitting thermal light is stellar imaging using a Hanbury-Brown-Twiss (HBT) intensity-correlation interferometer \cite{HBT}, where the maximum visibility achievable is limited by $\frac{1}{3}$.

The development of ghost imaging (GI) offers an intriguing optical technique to acquire the object's transverse transmittance pattern by means of photocurrent correlation measurements. The unique features of GI are that an image of the object is reconstructed by correlating the intensities of two spatially correlated beams. One of the beam illuminates the object and is detected by a bucket detector which has no spatial resolution. The
other reference beam undergoes only free-space diffraction before impinging on a scanning pinhole detector or a CCD camera with high spatial resolution. The first GI demonstration \cite{todd} explored entangled paired photons generated from spontaneous parametric down conversion together with photon-counting bucket and pinhole detectors more than a decade ago. Subsequent realizations with classical and especially (pseudo-)thermal light sources \cite{boyd2,boyd,han,valencia,gatti,scarcelli,wu,zhu,wang,wu10} triggered ongoing effort on applying GI to remote sensing applications \cite{meyers}. Unquestionably, the visibility of thermal-light GI in those experiments can never bypass the standard limit $\frac{1}{3}$. However, whether the nature of pseudothermal GI can be interpreted as classical intensity correlations \cite{gatti2,shapiro1,zhu2} or is fundamentally a
two-photon interference effect \cite{scarcelli,scarcelli2} is still under debate. Nonetheless, using GI for practical applications has attracted considerable attention in the community.

Recently, Shapiro proposed a modified version of thermal GI, called computational GI (CGI), in which the spatial intensity distribution measured in the reference beam is computed offline instead \cite{shapiro2}. CGI differs from previous thermal GI by replacing the rotating ground glass (RGG) with a spatial light modulator. The image is obtained by correlating the calculated field patterns with the measured intensities at the object arm. This CGI technique has been confirmed by two recent experiments \cite{silberberg,duran}. To reduce the burden of the computation in the virtual reference arm, the demonstrations of compressive GI \cite{duran,katz} provide a way by utilizing prior knowledge on the object for reducing the number of acquired measurements, without (significantly) sacrificing the image quality. Although these achievements are impressive, the image formation still fully relies on the precise \textit{measurements} from both sides, the test arm and the reference arm, no more beyond the frame of previous thermal GI \cite{boyd2,boyd,han,valencia,gatti,scarcelli,wu,zhu,wang,wu10}.

Very recently, a different but interesting observation on thermal GI was made in the Wu's group \cite{wu22}. Their experimental setup [see Fig. \ref{fig:Fig1}(b)] was similar as others \cite{valencia,gatti,scarcelli,wu,wu10} and the data collections as well, except for processing the data to form the image. They found that the reconstruction of the image can be conditionally obtained through partial measurements from the reference arm (even it never traverses the object) by introducing a novel algorithm. This finding is significant and differs from the image formation in conventional GI in the sense that through the algorithm, the images of the object can be \textit{computed} from the reference arm. They also found that the constructed images can be either positive or negative. We notice that a negative image was constructed in a lately reported experiment \cite{wang2}. However, the physics of forming such a negative image is fundamentally different from the one observed by Luo and her colleagues \cite{wu22}. Consistent with previous results in the literature, no image is observable in their experiment by applying all the measurements in the reference arm.

Inspired by their experiment \cite{wu22}, here we wish to provide a theoretical description on the experiment. Our theory not only offers a physical explanation on their findings, but also allows an in-depth understanding of thermal GI. In particular, we theoretically find that the method (or algorithm) discovered by Luo and her colleagues \cite{wu22} would also allow to construct images with the visibility arbitrarily close to unity, well beyond $\frac{1}{3}$, but with the sacrifice of the image quality. This leads us to give an interesting interpretation on the algorithm. That is, it may act as a ``bandpass filter" to subtract the DC background in thermal GI and thus enhances the image contrast. Besides, we further argue that, as an inverse problem, it would be possible to retrieve ``partial" image of the object only using the measurements from the reference arm conditioned on the number of objects present in the test arm, from the statistical viewpoint. Here ``partial" means that one can never precisely (or with 100$\%$ confidence) predict what the object is; and ``the number of objects" concerns how many objects are used for imaging during the data collection. Since the image formation only uses a bit information (akin to Yes or No) from the test arm, at first view, one might mistakenly draw a conclusion that the test (i.e. object) arm could be fully removed from the setup and the image was still achievable. In Sec. II, we will resolve this puzzle by emphasizing that, to form the \textit{real} image of the object with 100$\%$ confidence, spatial correlation of thermal light is indispensable, as required in all thermal GIs. Last, we emphasize that albeit the images are formed by mainly using the data from the reference arm, the process belongs to the second-order correlation measurement. In this paper, we take the thermal GI as an example, but the algorithm described here can be extended to thermal ghost interference \cite{scarcelli3,zhai} as well as GI and ghost interference with nonclassical light \cite{sue,wen} in the high-gain case. The binary operation for decoding the spatial correlation introduced here provides an alternative way for image formation. Based upon these findings ascribed above, we anticipate that our work will be useful for practical applications of GI and ghost interference.

\section{A Novel Algorithm for Ghost Imaging}
For completeness, we will first give a brief review on conventional thermal-light GI, then move onto the new observation on the image construction by using a portion of samplings from the reference arm conditioned on the object side. The goal of this paper is to find the physics behind the Wu's experiment \cite{wu22} and to privide a reasonable and consistent picture. The examination of the visibility or image contrast also allows us theoretically to predict that the image processing discussed here outperforms the conventional one.

\subsection{Brief Review of Thermal Ghost Imaging}
\begin{figure}[tbp]
\includegraphics[scale=0.58]{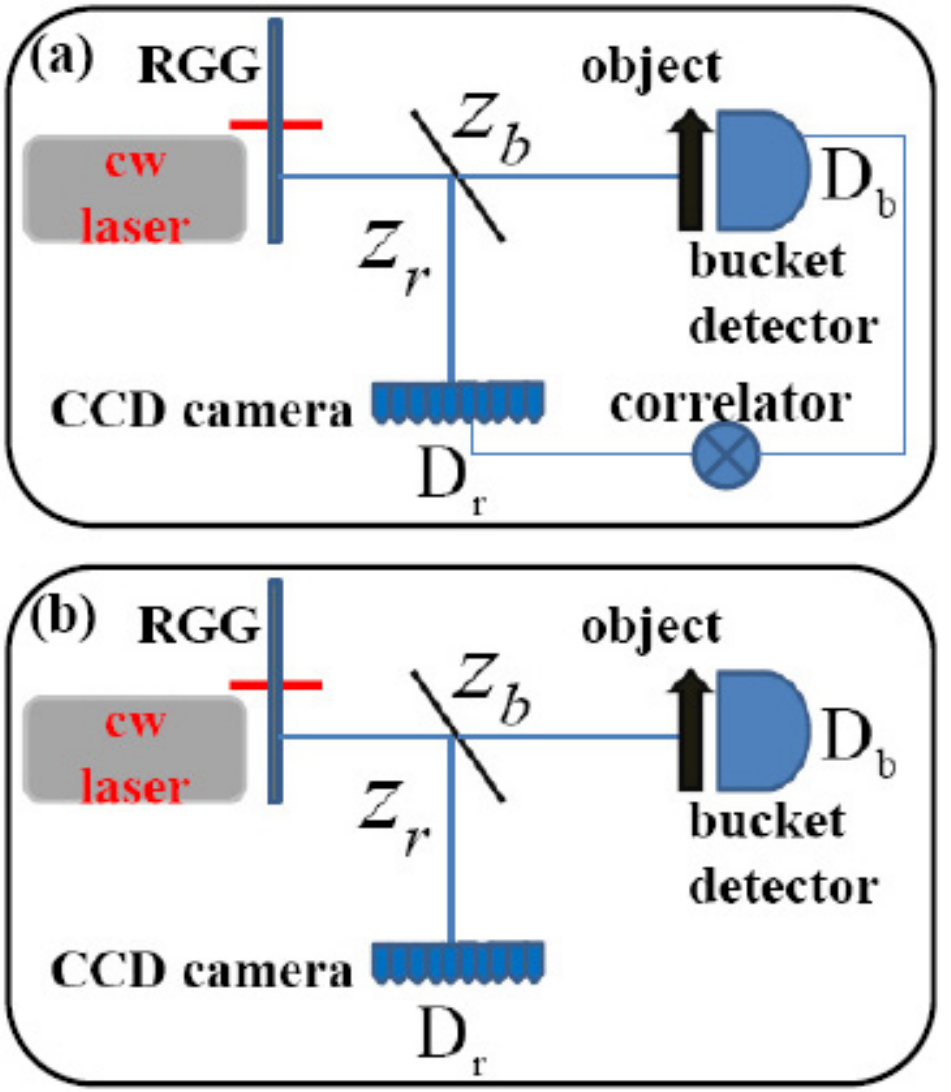}
\caption{(Color online) Conventional pseudothermal ghost imaging setup (a) and the experimental setup used in \cite{wu22} (b). RGG represents the rotating ground glass.}\label{fig:Fig1}
\end{figure}

To give an interpretation on the experiment done by Luo et al \cite{wu22}, we take the thermal lensless GI as an example to develop our theory. To ease the discussion, we begin with a brief review on the pseudothermal-light GI as schematically shown in Fig.~\ref{fig:Fig1}(a), using semiclassical photodetection theory. In the conventional pseudothermal GI, the test field $E_b$ generated by passing a cw laser through a slowly RGG and a 50-50 beam splitter illuminates an object and is detected by a bucket detector D$_b$. The reference field $E_r$ propagates freely towards
to a CCD camera D$_r$. The product of the photocurrents from D$_b$ and D$_r$, which is proportional to the second-order correlation function $G^{(2)}$, is time averaged to produce the ghost image of the object. It is convenient to write the positive-frequency part of the electromagnetic field $E_j(\vec{r}_j,t_j)$ ($j=b,r$) as a superposition of its longitudinal and transverse modes under the Fresnel paraxial approximation,
\begin{eqnarray}
E_j(\vec{\rho}_j,z_j,t_j)=\int{d}\vec{\kappa}{d}\omega\tilde{E}_j(\vec{\kappa},\omega)g_j(\vec{\kappa},\omega;\vec{\rho}_j,z_j)e^{-i\omega{t_j}},
\label{eq:Eq1}
\end{eqnarray}
where $\tilde{E}_j(\vec{\kappa},\omega)$ is the complex amplitude for the mode of angular frequency $\omega$ and transverse wavevector $\vec{\kappa}$. The Green's function, $g_j(\vec{\kappa},\omega;\vec{\rho}_j,z_j)$, which ascribes the propagation of each mode in space, can be evaluated as
\cite{rubin,goodman},
\begin{eqnarray}
g_b(\vec{\kappa},\omega;\vec{\rho}_b,z_b)&=&e^{i\frac{\omega{z}_b}{c}}\int{d}\vec{\rho}_oA(\vec{\rho}_o)e^{-i\frac{cz_b|\vec{\kappa}|^2}{2\omega}}
e^{i\vec{\rho}_o\cdot\vec{\kappa}},\label{eq:Eq2}\\
g_r(\vec{\kappa},\omega;\vec{\rho}_r,z_r)&=&e^{i\frac{\omega{z}_r}{c}}e^{-i\frac{cz_r|\vec{\kappa}|^2}{2\omega}}e^{i\vec{\rho}_r\cdot\vec{\kappa}}.
\label{eq:Eq3}
\end{eqnarray}
Here $A(\vec{\rho}_o)$ is the aperture function of the object, and $\vec{\rho}_{o}$ and $\vec{\rho}_{r}$ are, respectively, the transverse coordinates in the object and the CCD camera planes. $z_b$ is the distance from the output surface of the RGG to the object plane and $z_r$ is the length between the light source and D$_r$. The second-order correlation function $G^{(2)}$ is defined as \cite{glauber,mandel}
\begin{eqnarray}
G^{(2)}(\vec{\rho}_r)=\langle{E}^{\ast}_rE_rE^{\ast}_bE_b\rangle,\label{eq:Eq4}
\end{eqnarray}
where $\langle\cdot\rangle$ denotes the ensemble average. With use of Eqs. (\ref{eq:Eq1})-(\ref{eq:Eq3}), after some algebra, Eq. (\ref{eq:Eq4}) becomes \cite{han,valencia,gatti,scarcelli,wu,zhu,wang}
\begin{eqnarray}
G^{(2)}(\vec{\rho}_r)=G_0+G_0\left|\int{d}\vec{\rho}_oA(\vec{\rho}_o)\delta(\vec{\rho}_o-\vec{\rho}_r)\right|^2,\label{eq:Eq5}
\end{eqnarray}
or in terms of the normalized second-order correlation function $g^{(2)}$,
\begin{eqnarray}
g^{(2)}(\vec{\rho}_r)=1+\left|\int{d}\vec{\rho}_oA(\vec{\rho}_o)\delta(\vec{\rho}_o-\vec{\rho}_r)\right|^2,\label{eq:Eq6}
\end{eqnarray}
where $G_0$ is a constant. In the derivation of Eq. (\ref{eq:Eq5}), we have applied the lensless imaging condition $z_b=z_r$. In the literature, Eq. (\ref{eq:Eq5}) [as well as Eq. (\ref{eq:Eq6})] summarizes most of important properties of thermal GI. For example, the lensless ghost image is established through the intensity correlation measurement of the two beams, as confirmed from the second term on the right hand side of Eq. (\ref{eq:Eq5}), while the first term only contributes to a featureless background. The maximum visibility cannot pass the limit $\frac{1}{3}$, due to this background. Moreover, the point-to-point mapping relationship between the object plane and the imaging plane is evidenced by the Dirac $\delta$-function. Furthermore, no image is available if only looking at one arm.

\subsection{Conditioned Image Formation from the Reference Arm}
We turn our attention now to the experimental setup employed in Ref. \cite{wu22}, see Fig. \ref{fig:Fig1}(b). In their experiment, they recorded the data in both paths same as the conventional thermal GI experiments, see Refs. \cite{wu,wu10}. The major difference arises from the data processing to obtain the image. In the experiment performed by the Wu's group, they introduced a novel algorithm (which will be discussed shortly) to conditionally produce the image from the reference arm instead of the traditional way by simply correlating two photocurrents through the correlator. Their experimental demonstration indicated that images can be constructed with partial reference films by conditioning on a bit information from the object arm. The conditional image formation from the reference arm led us to rethink the imaging process of thermal or chaotic GI. We are interested in understanding the physics behind, in particular, the mechanism that allows to conditionally retrieve the object's information using the films recorded by the reference CCD camera. Before proceeding the discussion, we emphasize that the experiment is still a two-arm experiment. Although the images are mainly formed with use of the reference measurements, the bit information from the object arm is necessary and indispensable.

To attain definite answers to those questions, let us consider the following experiment [see Fig. 1(b)] with total $N$ realizations done in both bucket and reference detectors. Conventional GI and CGI use precise correlation between the two arms. In the experiment done by Luo and her coworkers \cite{wu22}, however, they showed that the image can be conditionally formed with a portion of reference samplings by applying a novel algorithm (explained below). Such an interesting observation introduces a new way of detecting the correlation between two sides. To show quantitatively whether it is possible to conditionally form the image of the object utilizing the data recorded in the reference arm, we begin with discretizing the second-order correlation function (\ref{eq:Eq6}) in terms of these $N$ samplings. The spatial distribution of the object $A(\vec{\rho}_o)$ appearing in Eq. (\ref{eq:Eq6}) can be recovered through the following linear operation:
\begin{eqnarray}
1+|A(\vec{\rho}_o)|^2=\frac{1}{N}\sum^{N}_{j=1}\frac{B_j}{\langle{B}\rangle}\frac{I_j(\vec{\rho}_r)}{\langle{I(\vec{\rho}_r)\rangle}},\label{eq:Eq7}
\end{eqnarray}
where $\langle{B}\rangle\equiv\frac{1}{N}\sum^{N}_{j=1}B_j$ represents the average value of the measured intensities $\{B_j\}$ over total $N$ realizations in the object arm, $\langle{I(\vec{\rho}_r)}\rangle\equiv\frac{1}{N}\sum^{N}_{j=1}I_j(\vec{\rho}_r)$ stands for the average intensity distribution over $N$ realizations in the reference arm, and $I_j(\vec{\rho}_r)$ denotes the intensity distribution recorded at spatial point D$_r$ of the CCD camera plane while its twin party $B_j$ at the bucket detector D$_b$. Equation (\ref{eq:Eq7}) clearly implies that in conventional thermal GI, the image is constructed experimentally by a linear superposition of the intensities $I_j(\vec{\rho})$ with the weights $B_j$ \cite{boyd2,boyd,valencia,gatti,scarcelli,wu,wang,wu10,silberberg,duran}. From Eq. (\ref{eq:Eq7}) we note that the intensity distribution $I_j(\vec{\rho}_r)$ also contains a total averaged constant DC background in each individual frame. It is easy to verify that utilizing all measurements as in conventional GI, it is difficult to obtain a negative image from Eq. (\ref{eq:Eq7}). Yet, Eq. (\ref{eq:Eq7}) already implies that the spatial distribution of the object, $|A(\vec{\rho}_o)|$, can be, in fact, conditionally retrieved from the reference arm by introducing the following algorithm
\begin{eqnarray}
\{
\begin{array}{ll}
+1, & \mathrm{if} \;B_j-\langle{B}\rangle>0;\\
-1, & \mathrm{if}\;B_j-\langle{B}\rangle<0.
\end{array}\label{eq:Eq8}
\end{eqnarray}
To see how this works, we rewrite Eq. (\ref{eq:Eq7}) as
\begin{eqnarray}
|A(\vec{\rho}_o)|^2=\frac{1}{N}\sum^{N}_{j=1}\left(\frac{B_j}{\langle{B}\rangle}-1\right)\frac{I_j(\vec{\rho}_r)}
{\langle{I(\vec{\rho}_r)}},\label{eq:Eq71}
\end{eqnarray}
by using the identity $1=\frac{1}{N}\sum_{j=1}^{N}[I_j(\vec{\rho}_r)/\langle{I(\vec{\rho}_r)}\rangle]$. It is interesting to note that the rule (\ref{eq:Eq8}) introduces a binary operation for data processing. Moreover, it would allow one to obtain either a positive or negative image of the object with partial measurements from the reference arm, conditioned on the bucket side. To verify this, it is a simple matter of fact by substituting (\ref{eq:Eq8}) into Eq. (\ref{eq:Eq71}), which yields
\begin{eqnarray}
\pm|A(\vec{\rho}_o)|^2\cong\frac{1}{M}\sum^{M<N}_{j=1}\mathrm{sgn}\left(\frac{B_j}{\langle{B}\rangle}-1\right)\frac{I_j(\vec{\rho}_r)}
{\langle{I(\vec{\rho}_r)}\rangle},\label{eq:Eq9}
\end{eqnarray}
with
\begin{eqnarray}
\mathrm{sgn}(x)=\{
\begin{array}{lll}
1, & \mathrm{if}\;x>0;\\
0, & \mathrm{if}\; x=0;\\
-1, &\mathrm{if}\;x<0.
\end{array}\nonumber
\end{eqnarray}

\begin{figure}[tbp]
\includegraphics[scale=0.58]{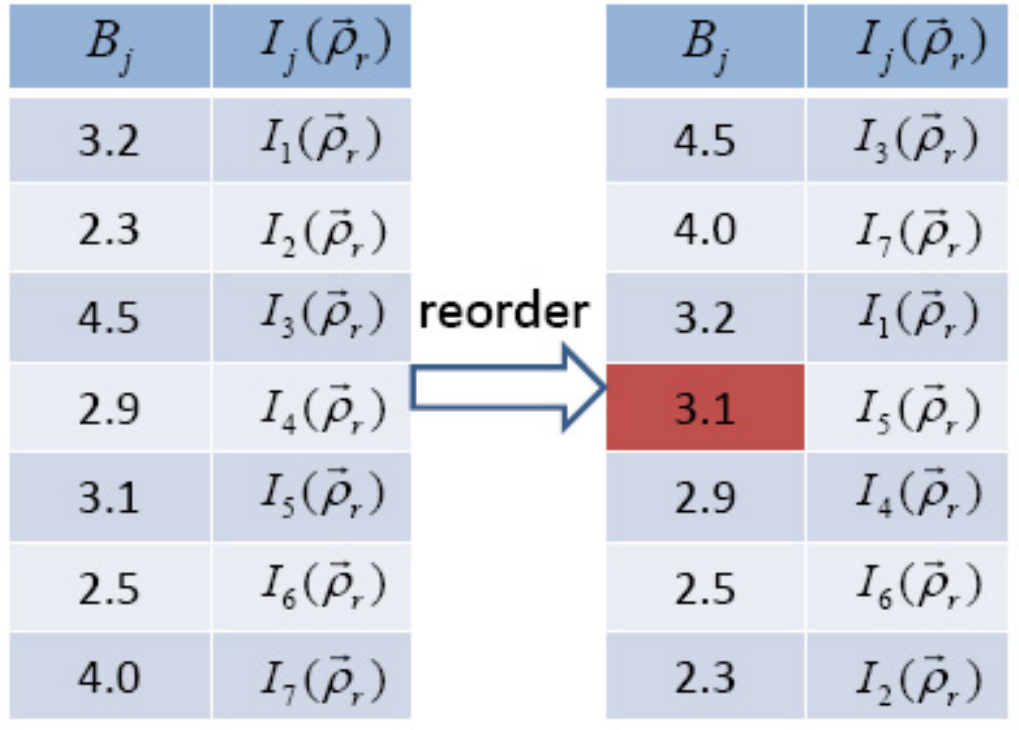}
\caption{(Color online) Illustration of the algorithm described in the context. By reordering the data, positive (or negative) images can be formed, for example, by summing some of $I_1$, $I_3$, and $I_7$ (or $I_2$, $I_4$, and $I_6$).}\label{fig:Fig2}
\end{figure}

Before proceeding the discussions, few remarks on Eqs. (\ref{eq:Eq8}) and (\ref{eq:Eq9}) are in order: (a) Whether the constructed image is positive or negative is fully determined by the sign of $\frac{1}{M}\sum^{M<N}_{j=1}\mathrm{sgn}(B_j-\langle{B}\rangle)$ with $M$ realizations. (b) By summing all measurements in the reference arm, Eq. (\ref{eq:Eq9}) statistically reduces to a featureless intensity distribution, which coincides with previous conclusions. (c) The averaged intensity value $\langle{B}\rangle$ in the test arm plays the role of being a reference number or a pointer to evaluate the correlation. It roughly divides the data into three blocks: one block fluctuates above $\langle{B}\rangle$; the second below $\langle{B}\rangle$; and the third around $\langle{B}\rangle$. In the experiment, in fact, this value can be chosen from one of the middle values in the third block from the bucket detector. This is exactly implemented by Luo and her coworkers \cite{wu22} in their experiment. That is, instead of really calculating the average intensity $\langle{B}\rangle$, one first needs to \textit{reorder} the intensities recorded by D$_b$, say, from the largest to the smallest; and the partner intensity patterns recorded by D$_r$ are then accordingly re-sorted. Whether the formed image is positive or negative is simply dependent on $I_j$, whose partners $B_j$ are above or below the middle $B_\mathrm{mid}$. In Fig. \ref{fig:Fig2}, we used a flow chart to illustrate the method. Suppose in an experiment, 7 measurements were performed in each arm. By reordering $B_j$ and setting $I_5$ as the reference intensity $\langle{B}\rangle$, the positive images can be constructed by summing $I_1$, $I_3$, and $I_7$; while the negative ones can be formed by adding $I_2$, $I_4$, and $I_6$. To be more concrete, in Fig. 3 we have looked at a double-slit ghost imaging with thermal light, where about 50,000 films have been recorded in the reference arm. Figure 3(a) gives the ghost image of the double slit through the conventional $G^{(2)}$ method. The images shown in Figs. 3(b1)-(c2) are obtained through the algorithm described above. Figures 3(b1) and (b2) give examples of negative images where $\sim$26,000 and $\sim$1,400 films from the reference arm have been used respectively. Figures 3(c1) and (c2) give examples of positive images where $\sim$18,000 and $\sim$1,600 reference films have been chosen. (d) The algorithm described here may be useful for the compressive GI \cite{silberberg,duran}. Since above or below $\langle{B}\rangle$ the formed images have a relative high visibility (explained below), these samplings are more helpful for image constructions than those around $\langle{B}\rangle$. Of course, mixing the data above and below $\langle{B}\rangle$ together results in the reduction of the visibility, in comparison with the data just either above or below $\langle{B}\rangle$. From this point of view, compressive measurements are allowable for reference data whose partner is above or below the averaged value $\langle{B}\rangle$. (e) The resolution of the image may be affected by the introduced algorithm (\ref{eq:Eq8}). A concrete discussion on this issue is beyond the scope of the current paper, and will be addressed somewhere else.
\begin{figure}[tbp]
\includegraphics[scale=0.58]{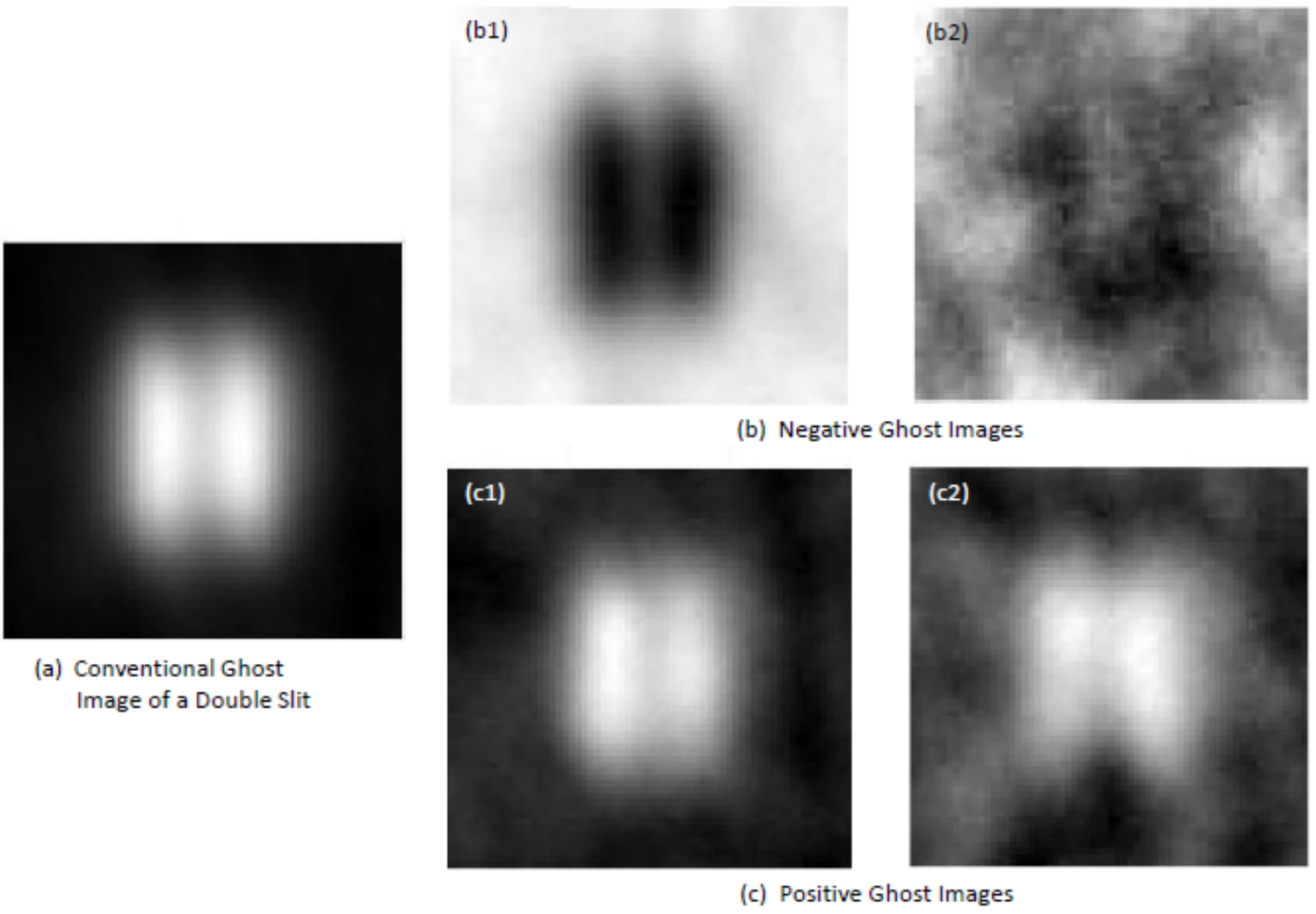}
\caption{A double-slit thermal ghost imaging experiment. (a) The image is formed through the conventional second-order correlation measurements where $\sim$50,000 measurements have been recorded in each arm. (b1) and (b2) give examples of negative images formed by applying the algorithm (8). (c1) and (c2) illustrate the cases of positive image formations with use of the algorithm (8).}\label{fig:Fig3}
\end{figure}

Finally, let us look at the issue of the visibility (or image contrast). Interestingly, the algorithm introduced in Eq. (\ref{eq:Eq8}) could allow to construct images with visibility well beyond $\frac{1}{3}$. This conclusion seems apparently contradict with our common knowledge learnt from the HBT experiment \cite{HBT}. How could this be true? To resolve the contradiction, we rewrite Eq. (\ref{eq:Eq7}) as
\begin{eqnarray}
|A(\vec{\rho}_o)|^2=\frac{1}{N}\sum^{N}_{j=1}\left(\frac{B_j}{\langle{B}\rangle}-1\right)\left[\frac{I_j(\vec{\rho}_o)}{\langle{I(\vec{\rho}_o)
\rangle}}-1\right].\label{eq:Eq10}
\end{eqnarray}
Again, we have applied the identities $1=\frac{1}{N}\sum_{j=1}^{N}[I_j(\vec{\rho}_r)/\langle{I(\vec{\rho}_r)}\rangle]$ and $1=\frac{1}{N}\sum_{j=1}^N(B_j/\langle{B}\rangle)$. Note that Eq. (\ref{eq:Eq10}) is a reformulation of the intensity correlation fluctuation in the HBT measurements. Moreover, Eq. (\ref{eq:Eq10}) states that if the runs $B_j$ are greater (or lower) than $\langle{B}\rangle$, their twin parties $I_j(\vec{\rho}_o)$ are also bigger (or lower) than $\langle{I(\vec{\rho}_o)}\rangle$. In light of college optics, we know that the visibility of the image is defined as
\begin{eqnarray}
V=\pm\frac{I_{\mathrm{max}}-I_{\mathrm{min}}}{I_{\mathrm{max}}+I_{\mathrm{min}}},\label{eq:Eq11}
\end{eqnarray}
where $I_{\mathrm{max}}$ and $I_{\mathrm{min}}$ are the maximum and minimum intensities, respectively, and $\pm$ are for positive and negative images. It is known that the fluctuation of thermal light with average intensity $\bar{I}$ is equal to $\bar{I}$ \cite{mandel}. Alternatively, the instantaneous intensities far away from $\bar{I}$ have much larger fluctuations than $\bar{I}$ but with smaller probability of production. If we now choose $M$ measurements $I_j(\vec{\rho}_o)$, whose correlated parties $B_j$ are far above or below $\langle{B}\rangle$ in the first or second block mentioned above by satisfying $M\ll{N}$, it is not difficult to verify that the visibility (\ref{eq:Eq11}) of the formed image from the reference arm can be made arbitrarily close to $\pm1$. For the positive image, we have
\begin{eqnarray}
V=\frac{\sum_{j=1}^{M}\frac{I_j(\vec{\rho}_o)}{\langle{I(\vec{\rho}_o)}\rangle}\Big|_{\mathrm{max}}-M}{\sum_{j=1}^{M}\frac{I_j(\vec{\rho}_o)}
{\langle{I(\vec{\rho}_o)}\rangle}\Big|_{\mathrm{max}}+M}<1,\label{eq:Eq12}
\end{eqnarray}
by noticing the fact $I_j(\vec{\rho}_o)/\langle{I(\vec{\rho}_o)}\rangle\gg1$. However, for the negative image we have
\begin{eqnarray}
-V=\frac{M-\sum_{j=1}^{M}\frac{I_j(\vec{\rho}_o)}{\langle{I(\vec{\rho}_o)}\rangle}\Big|_{\mathrm{min}}}{M+\sum_{j=1}^{M}\frac{I_j(\vec{\rho}_o)}
{\langle{I(\vec{\rho}_o)}\rangle}\Big|_{\mathrm{min}}}<1,\label{eq:Eq13}
\end{eqnarray}
with $I_j(\vec{\rho}_o)/\langle{I(\vec{\rho}_o)}\rangle\ll1$. Such a high visibility is comparable with that obtained with entangled photon pairs \cite{todd}, where the visibility of almost 100$\%$ is usually considered as a \textit{signature} of using biphotons without background subtraction. The price of achieving a high visibility shown in Eqs. (\ref{eq:Eq12}) and (\ref{eq:Eq13}) is paid by using only a small portion of samplings but discarding the majority. Consequently, this may result in the deduction of the image quality, e.g., the spatial resolution. So, a better quality would compromise with a reduced visibility \cite{gatti}. With use of more reference data, the contrast of precise image of the object will eventually drop to the ideal limit $1/3$, in agreement with textbook knowledge. From Eqs. (\ref{eq:Eq12}) and (\ref{eq:Eq13}), the present method introduces a way to \textit{eliminate} the constant background from the $G^{(2)}$ or $g^{(2)}$ measurement [reference to Eq. (\ref{eq:Eq5}) or Eq. (\ref{eq:Eq6})]. That is, the intensities that well distribute around the average value more contribute to the background and hence result in a lower visibility. From this point of view, the algorithm discussed in this paper behaves as an imaginary \textit{DC blocker}, even which was not really present in the detection system. It is also obvious that mixing measurements above and below the (average) reference pointer will decrease the visibility substantially. Last, the discussion presented here shows another way to decipher the spatial correlation of twin beams and shines a new light for understanding the amplitude fluctuation of stochastic process, especially, thermal light.

\subsection{Further Discussions}

Before ending the discussions, we wish to add further few remarks. First of all, the algorithm presented here is also applicable to thermal ghost interference \cite{scarcelli3,zhai} as well as ghost imaging and ghost interference with optical parametric amplifier \cite{sue} or four-wave mixing \cite{wen}, because of their comparability with thermal light. In comparison with thermal light, for nonclassical light the Gaussian thin lens equation is required to realize the point-to-point mapping between the object plane and the imaging plane. Secondly, the topic presented here may be useful for analyzing some features in optical encryption \cite{duran}, such as key compressibility and vulnerability to eavesdropping. Thirdly, as an inverse problem, one might speculate that with the help of a super-fast computer, the image of the object could be ultimately identified from the reference arm simply through evaluating all kinds of permutations and combinations. In other words, one would expect to recover the spatial correlation through this numerical random-data processing. However, one problem arises against such an apparent paradox. That is, one at least needs to know among the analyzed data from the reference arm, how many objects have been imaged in the test arm. If the evaluated reference data corresponds to that only one object is placed in the object arm, it may be possible to deduce its spatial distribution only with the measurements from the reference arm, but with less than 100$\%$ confidence. That means partial but not full knowledge about the object's profile could be read out in the case. The reduced confidence stems from the lack of knowledge from the object side, i.e., the lack of the precise spatial
correlation of thermal light.

To make the point clear, let us imagine that in a thermal GI experiment with total $P$ measurements, due to some unknown reasons one completely loses the data from the bucket detector. The only information which we have in mind is that only one still object is present in the object arm during the $P$ measurements. The question now is: Could we still use the films recorded in the reference CCD to form the image? The answer to this question is Yes. Recall Eq. (\ref{eq:Eq9}) which tells that if a conditionally formed image with partial $Q$ reference measurements is positive, the image with rest $P-Q$ measurements would be negative. Here $Q$ is chosen to be either greater or lower than $\frac{P}{2}$. The shapes of two images look the same (similar as the complementarity principle in college optics books). Due to lack the data from the object side the formed images may have poor visibility in most time. If the visibility is too poor to recognize what the image looks like, one then exchanges, say, $T$ measurements from previous $Q$ and $P-Q$ measurements and repeats the procedure stated above, untill one can discriminate what the object is from the series of computed images with acceptable visibility. Here $T$ should be bigger than 1 but less than minimum($Q,\frac{P}{2}$).

Based upon this finding, our protocol admits \textit{numerical image formation} with a single detector from a statistical viewpoint. Caution should be urged in the statement, that is, the condition is only one object in the test arm during the data collection. Here, we emphasize again that to have a ghost image of the object with 100$\%$ confidence as observed in Refs. \cite{valencia,gatti,scarcelli,wu,wang,wu10,meyers}, the spatial correlation of light is indispensable and the measurements from both arms are necessary. It is certainly true that by sorting a random set of noise pictures to get an image, but one can get any image.  From this aspect, the topic presented in this paper shares a link with random data processing. Along with the increase of the number of objects, it turns out to be a more difficult and eventually impossible task to discriminate them one by one from the noise. Based on this point, the algorithm may further reveal its advantages over conventional full correlation measurements in multi-channel thermal GI setups. For instance, simultaneous performing thermal GI with two or more than two objects in the same runs. Further discussions on this issue might be presented elsewhere.

In comparison with CGI, the current scheme conditionally forms the image with use of partial measurements from the reference arm; while CGI forms the image precisely using all the measurements from both the bucket side and the computed reference side, same as conventional GI. Both schemes obtain the image through introducing an algorithm. However, the present method is more useful if the measurements from the object arm were lost.

\section{Summary}
In summary, we provide a simple theory to explain a recent thermal GI experiment \cite{wu22} in which either a positive or a negative image can be constructed with only partial measurements from the reference arm by applying a novel algorithm (\ref{eq:Eq8}). In contrast with conventional GI, the algorithm offers a new way to decipher the spatial correlation of thermal light (by sacrificing the quality). Particularly, we theoretically predict that the scheme used by Luo and her colleagues \cite{wu22} can outperform conventional GI in that the images can have a visibility higher than $\frac{1}{3}$, and ideally approaching the unity. Unity visibility used to be thought of a signature of using entangled photons, where no background subtraction is required. We further show that it is possible to numerically form the image by sampling the reference data, if the data from the object side is lost. The condition is that only one still object is present in the test arm for the data of interest. The results presented here not only are consistent with all previous research on GI, but also open a new way for the GI formation. Although the origin of formed images seems from the \textit{first-order correlation function}, in fact they originate from the \textit{second-order correlation function} [see Eq. (\ref{eq:Eq9})] in the sense that a bit information from the object arm is utilized. Although the presented discussion is made on the spatial correlation of thermal light, it is expected that the described algorithm would be applicable to the temporal domain (e.g. temporal ghost imaging with thermal light). Finally, the new features and properties discovered in this paper may be useful for turning GI into reality.

\section{Acknowledgements}
We are grateful to Morton H. Rubin, Giuliao Scarcelli, Yan-Hua Zhai, and Yoon-Ho Kim for their illuminating discussions and critical comments on the topic presented here. We specially thank Kai-Hong Luo and Ling-An Wu for letting him know their experimental observations on forming negative and positive images by introducing the algorithm at the early stage. This work was partially supported by the 111 Project B07026 (China). We also acknowledge the financial supports from the AI-TF New Faculty Grant and the NSERC Discovery Grant in Canada.

\end{document}